\documentclass[
reprint,
superscriptaddress,
 amsmath,amssymb,
 aps,
prb,
floatfix,
]{revtex4-1}
\usepackage{graphicx}% Include figure files
\usepackage{dcolumn}% Align table columns on decimal point
\usepackage{bm}% bold math

\usepackage{CJK}
\usepackage{times}
\usepackage{amsmath}
\usepackage{amssymb}
\usepackage[colorlinks,linkcolor=blue,citecolor=blue,urlcolor=blue,hyperindex,pdfstartview=FitH,plainpages=false]{hyperref}
\usepackage{float}

\newcommand {\TNS}{Ta$_2$NiSe$_5$}
\newcommand {\mJcm}{mJ/cm$^2$}

\begin{document}
\begin{CJK*}{GBK}{}
\title{Non-Coulomb strong electron-hole binding in \TNS~revealed by time- and angle-resolved photoemission spectroscopy}

\author{Tianwei Tang}
\affiliation{Key Laboratory of Artificial Structures and Quantum Control (Ministry of Education), Shenyang National Laboratory for Materials Science, School of Physics and Astronomy, Shanghai Jiao Tong University, Shanghai 200240, China}
\author{Hongyuan Wang}
\affiliation{School of Physical Science and Technology, ShanghaiTech University, Shanghai 201210, China}
\author{Shaofeng Duan}
\author{Yuanyuan Yang}
\author{Chaozhi Huang}
\affiliation{Key Laboratory of Artificial Structures and Quantum Control (Ministry of Education), Shenyang National Laboratory for Materials Science, School of Physics and Astronomy, Shanghai Jiao Tong University, Shanghai 200240, China}
\author{Yanfeng Guo}
\affiliation{School of Physical Science and Technology, ShanghaiTech University, Shanghai 201210, China}
\author{Dong Qian}
\author{Wentao Zhang}
\email{wentaozhang@sjtu.edu.cn}
\affiliation{Key Laboratory of Artificial Structures and Quantum Control (Ministry of Education), Shenyang National Laboratory for Materials Science, School of Physics and Astronomy, Shanghai Jiao Tong University, Shanghai 200240, China}
\date {\today}

\begin{abstract}

We reveal an ultrafast purely electronic phase transition in \TNS, which is a plausible excitonic insulator, after excited by an ultrafast infrared laser pulse. Specifically, the order parameter of the strong electron-hole binding shrinks with enhancing the pump pulse, and above a critical pump fluence, a photo-excited semimetallic state is experimentally identified with the absence of ultrafast structural transition. In addition, the bare valence and conduction bands and also the effective effective masses in \TNS~are determined. These findings and detailed analysis suggest a bare nonequilibrium semimetallic phase in \TNS~and the strong electron-hole binding cannot be exclusively driven by Coulomb interaction.
\end{abstract}

\maketitle
%\clearpage
\end{CJK*}

\section{\label{sec:level1}INTRODUCTION}

In narrow gaped semimetals or semiconductors the bound electron-hole pairs will create a clean energy gap when the binding energy exceeds the band gap, resulting an excitonic insulating phase\cite{Mott1961,Kozlov1965,Jerome1967}. The excitonic insulating transition is purely electronic origin, and the ground state is analogous to that of the BCS superconductivity but with a diagonal kind of order. It is hard to realize such a novel insulating phase in elements of semimetals such as As, Sb, and Bi because of the small electron-hole binding energies due to the high dielectric constants and small effective masses. Only in recent years, such novel electronic phase was proposed in strongly correlated systems, such as 1T-TiSe$_2$\cite{Cercellier2007}, TmSe$_{0.45}$Te$_{0.55}$\cite{Bronold2006,Bucher1991}, and \TNS\cite{Wakisaka2009,Seki2014}, and also in the interface of InAs/GaSb bilayers\cite{Du2017}. However, strong electron-hole binding induced charge density wave or structural phase transition usually happens at the same critical temperature, making it difficult to distinguish the excitonic insulating phase from those transitions experimentally. Experimental evidence of the purely electronic phase transition is necessary in isolating the excitonic insulating physics in materials.

\TNS, a layered compound stacked by weak van der Waals force, shows a second order phase transition at 326 K while the crystal structure transfers from a orthorhombic structure to a monoclinic phase at the same critical temperature\cite{Sunshine1985,DiSalvo1986,Seki2014,Lu2017}.
An energy gap of about 0.15 eV is developed below the transition temperature ($T_c$) from optical and also photoemission spectroscopies, and it evolves smoothly crossing the $T_c$\cite{Wakisaka2012,Seki2014}. It was generally believed to be an excitonic insulator due to strong Coulomb interaction between the Ni 3d-Se 4p hybridized hole and the Ta 5d electron with positively gaped bare valence and conduction bands at the $\Gamma$ point, favoring the BEC phase transition\cite{Bronold2006,Ihle2008,Wakisaka2009,Phan2010,Ejima2014,Sugimoto2016a,Werdehausen2018}, and only recently a possible semimetal picture is established\cite{Yamada2016,Sugimoto2018,Tanabe2018,Domon2018,Okazaki2018,Lee2019,Fukutani2019,Mazza2019}. As a strongly correlated material, there is lack of theory to calculate the bare dispersions in \TNS~precisely, and usually modeled bare valance and conduction bands are adopted to fit the occupied excitonic band dispersion to extract the order parameter and also the bare bands\cite{Seki2014}. A modeless method to determine the bare bands and also the effective masses is important in understanding the excitonic insulating physics in \TNS.

Time- and angle-resolved photoemission spectroscopy (TrARPES) is capable of tracking the ultrafast electronic dynamics after melting the electron-hole pairs by ultrafast photon excitations and then determining the bare dispersions. In this paper we will report an observation of the ultrafast purely electronic phase transition in \TNS~by trARPES. We find that laser pumping greatly reduces the electron-hole binding induced energy gap without an ultrafast structural transition. Above a critical fluence, bare conduction and valence bands with negative band gap at the $\Gamma$ point are clearly evidenced in the nonequilibrium spectra. With trARPES data of high momentum and energy resolutions and the mean field ground states of an excitonic insulator, the bare dispersions of both the electron and hole bands and the effective masses are solved from the experimental time-dependent non-equilibrium Bogoliubov dispersions. Detailed analysis suggests a bare semi-metallic state in \TNS~and the strong binding energy of the exciton cannot be a result of only Coulomb interaction.

%%%%%%%%%%%%%%%%%%%%%%%%%%%%%%%%%
%Fig1
%%%%%%%%%%%%%%%%%%%%%%%%%%%%%%%%%
\begin{figure*}%[H]
\centering\includegraphics[width=2\columnwidth]{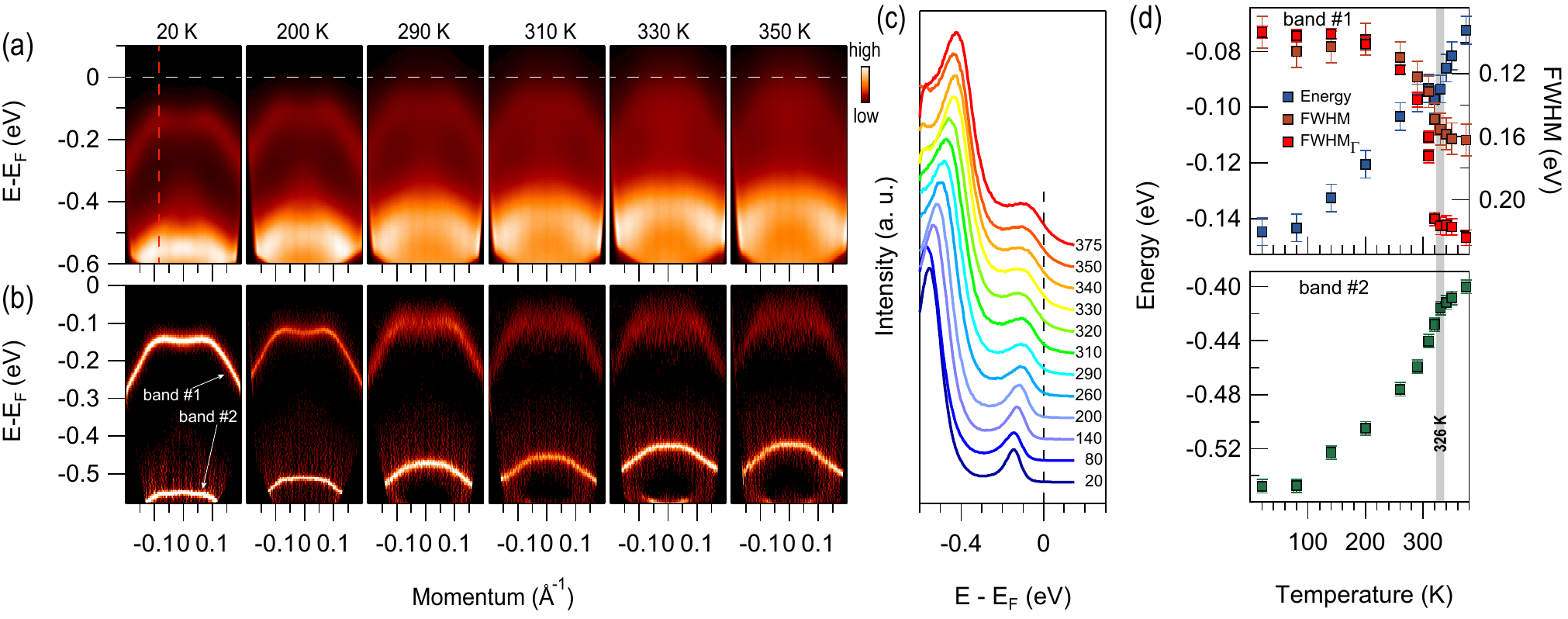}
\caption{
Temperature dependence of equilibrium photoemission spectra near the $\Gamma$ point along the Ta and Ni chains.
(a) Photoemission intensity as a function of energy and momentum at temperatures from 20 to 350 K.
(b) Corresponding curvature images of the spectra shown in panel (a).
(c) Temperature dependent energy distribution curve at the momenta of the band top (red dashed line cut in panel (a)).
(d) Energy gap of the bands \#1 and binding energy of band \#2 and the FWHM at the band top and also the FWHM at $\Gamma$ of the band \#1 as functions of temperature from panel (c).
}
\label{Fig1}
\end{figure*}

\section{\label{sec:level2}EXPERIMENTAL DETAILS}

In our trARPES experiments\cite{Yang2019}, an infrared pump laser pulse ($h\nu$=1.77 eV) with the pulse duration of 30 fs, repetition rate of 500 kHz, and the polarization perpendicular to the Ta and Ni chains drives the sample into a non-equilibrium state. An ultraviolet probe laser pulse (6.05 eV) subsequently photoemits electrons, which are captured by a hemispherical analyzer in an ARPES setup. The polarization of the probe beam is also perpendicular to the Ta and Ni chains. The overall time resolution is 130 fs and the energy resolution is 19 meV in the measurements. \TNS~single crystals were grown by the flux method and cleaved under ultrahigh vacuum condition with a pressure better than 3.5$\times$10$^{-11}$ Torr.

\section{\label{sec:level3}RESULTS AND DISCUSSION}

Figure~\ref{Fig1} shows the temperature dependence of the equilibrium photoemission spectra. Clear ``Mexican-hat'' dispersion (band \#1), which is a signature of band mixing of the conduction and valence bands due to strong electron-hole interaction\cite{Armitage2018}, can be evidenced at temperature from far below $T_c$ to above $T_c$ (panels (a) and panel (b)). In addition, the other high energy band (band \#2) with band top sitting at $\sim$ -0.53 eV and a faint dispersion with band top at $\sim$ -0.35 eV can be evidenced. Both the bands \#1 and \#2 shift to lower binding energies with heating the sample, as shown in the temperature dependent energy distribution curve (EDC) at the momenta -0.095 $\AA^{-1}$ where the band is closest to the Fermi energy (panel (c)). Far below the transition temperature at 20 K, the band \#1 shows an energy gap of $\Delta$ $\sim$ 0.15 eV, which is the excitonic insulating order parameter and the $2\Delta$ is consistent with the tunneling spectroscopic results\cite{Lee2019}. The energy gap shrinks smoothly to 0.08 eV at 375 K (panel (d)), agreeing with previous reports\cite{Seki2014,Wakisaka2012,Larkin2017}. Moreover, the scattering rate of the electron-hole mixing states shows an abrupt drop from $\sim$0.22 eV at $\Gamma$ ($\sim$ 0.16 eV at the band top) above $T_c$ to $\sim$ 0.1 eV at low temperature, consistent with an optical study\cite{Larkin2017}. Since the electronic states near the gap edge are analogous to the Bogoliubov dispersion in BCS theory, in which the quasiparticle lifetime (scattering rate) is related to the condensation of the Cooper pairs, the abrupt drop of scattering rate near $T_c$ here is a signature of the condensation of excitons. The absence of a temperature scale at $T_c$ in energy gap and the enhancement of the quasiparticle lifetime below $T_c$ is a signature of BEC, analogous to the precursor pairing scenario in cuprate high-temperature superconductivity\cite{Emery1995a}. The energy of the band \#2 drops from -0.40 eV at 375 K to -0.54 eV at 20 K, and exhibits a transition at T$_c$, which might be a result of structural transition and less screening effect below T$_c$.

%%%%%%%%%%%%%%%%%%%%%%%%%%%%%%%%%
%Fig2
%%%%%%%%%%%%%%%%%%%%%%%%%%%%%%%%%
\begin{figure}
\centering\includegraphics[width=1\columnwidth]{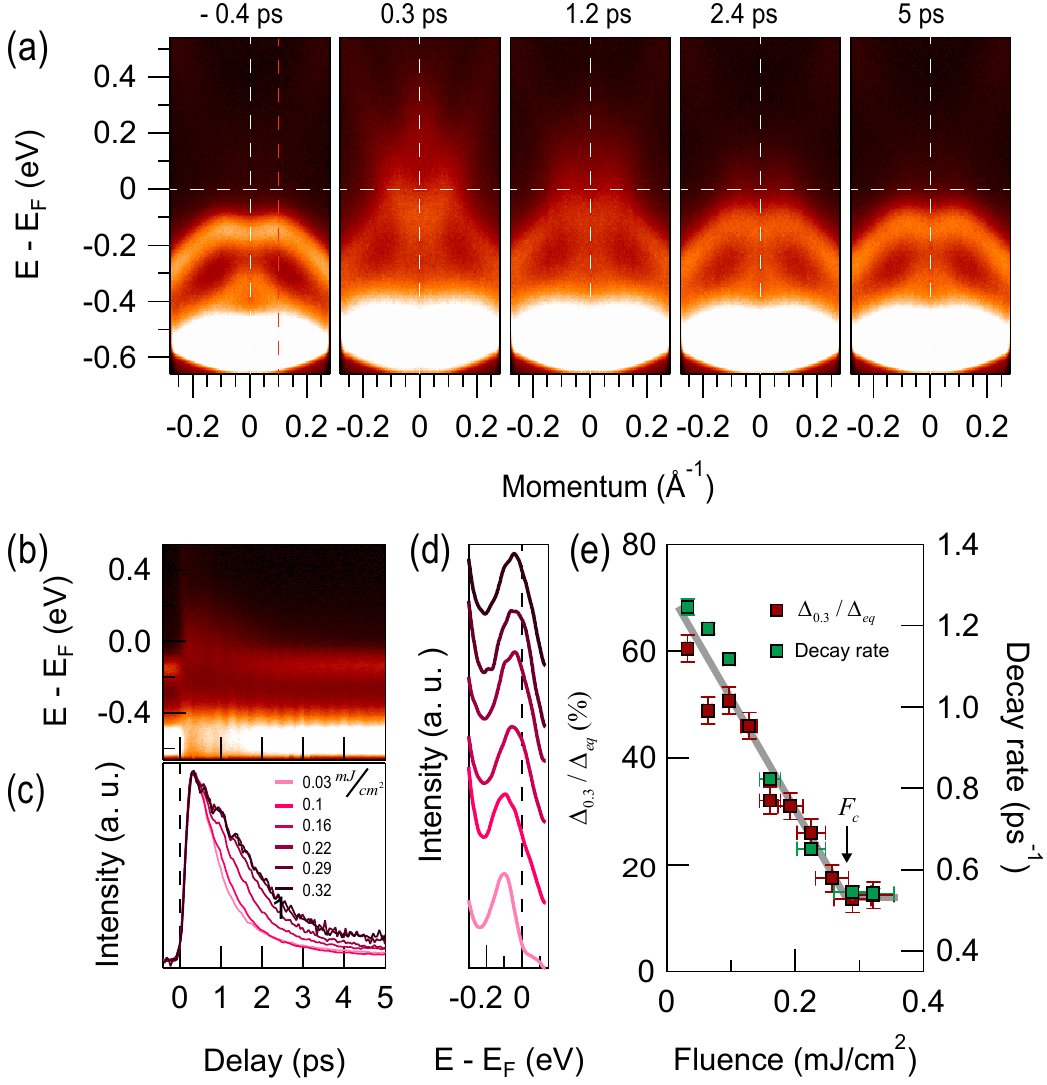}
\caption{
Pump-induced energy gap change at 30 K.
(a) Time-resolved photoemission spectra with the pump fluence at 0.19 \mJcm.
(b) Photoemission intensity as a function of energy and delay time at the momentum of band top (red dashed line cut in panel (a)).
(c) Intensity of pump-induced non-equilibrium electrons integrated in the energy window 0.1 eV above the Fermi energy as functions of delay time at different pump fluences. All the curves are normalized to the same height.
(d) Energy distribution curves at the momenta of the band top and delay time 0.3 ps for the same pump fluences.
(e) The pump induced change of energy gap from the panel (d) and the decay rate of non-equilibrium electrons as functions of pump fluence.
}
\label{Fig2}
\end{figure}

In Fig.~\ref{Fig2}  we use ultrafast laser pulses to excite the sample to reach higher electronic temperature. With a pump fluence of 0.19 \mJcm, the energy gap at 0.3 ps is greatly reduced and the nonequilibrium spectra clearly shows two branches of the mixed conduction and valence bands (panel (a)), which are still gaped as analyzed later. The upper branch is still distinguishable at 1.2 ps. With a repetition rate at 500 kHz and a base temperature at 30 K, little laser induced heating effect is noticed even for the highest pump fluence used (supplemental discussion \#1 and Fig. S1)\cite{SM}. The electronic states recover slowly but not fully to its equilibrium value for the longest delay time (8 ps) measured in this experiment. In addition, the time-resolved intensity at the momenta of the band top (panel (b)) clearly shows that the bands \#1 and \#2 oscillate in energy at some frequencies which will be discussed later on.
Consistent with reported by Mor et al. \cite{Mor2017}, photo-voltage effect is absent in this material (supplemental discussion \#2 and Fig. S2\cite{SM}).
%%,  and the pump-induced band shift is different from that in the surface doping experiment reported by Fukutani et al.\cite{Fukutani2019}
The observed pump-induced energy bands shifts cannot be a result of photo-induced Stark effect, which usually happens only in a short time scale and shifts the band to higher binding energy\cite{Sie2015,Chen2019}.

We find that the with higher pump fluence the intensity of non-equilibrium electrons recovers slower and the initial recovery rate (supplemental discussion \#3\cite{SM}) saturates above a critical pump fluence $\sim$ 0.29 \mJcm~($F_c$) (panels (c) and (e) in Fig.~\ref{Fig2}). The slower recovering rate at higher pump fluence is possibly due to smaller nonequilibrium gap and stronger screening effect with more excited charge carriers. It is different from pumping a superconducting system that the recombination of non-equilibrium electrons is faster with more excited non-equilibrium electrons, captured by the bimolecular model\cite{Smallwood2012}. With pumping harder, the low energy peak moves closer towards the Fermi level (panel (d)). The energy gap is dramatically suppressed to less than 15\% of the whole gap and saturates at the same pump fluence of $\sim$ 0.29 \mJcm~ as that of the non-equilibrium quasiparticle decay rate (panel (e)).
We note that the transient electronic temperature at the critical pump fluence is nearly 2000 K (supplemental Fig. S4\cite{SM}).
%Theoretical analysis suggests that the pump induced energy gap closure is primarily due to energy transfer from the photoexcited carriers to the exciton condensate\cite{Golez2016}.
The residue energy gap of 15\% is possibly attributed to the band avoided crossings at generic points in Brillouin zone. These results indicate that with pump above the $F_c$ the system undergoes an ultrafast electronic transition from an excitonic insulating state to a semimetalic state. The reported enhancement of the energy gap by Mor et al.\cite{Mor2017} is absent in our measurements with much lower equilibrium temperature and different pump wavelength.

%%%%%%%%%%%%%%%%%%%%%%%%%%%%%%%%%
%Fig3
%%%%%%%%%%%%%%%%%%%%%%%%%%%%%%%%%
\begin{figure}
\centering\includegraphics[width=1\columnwidth]{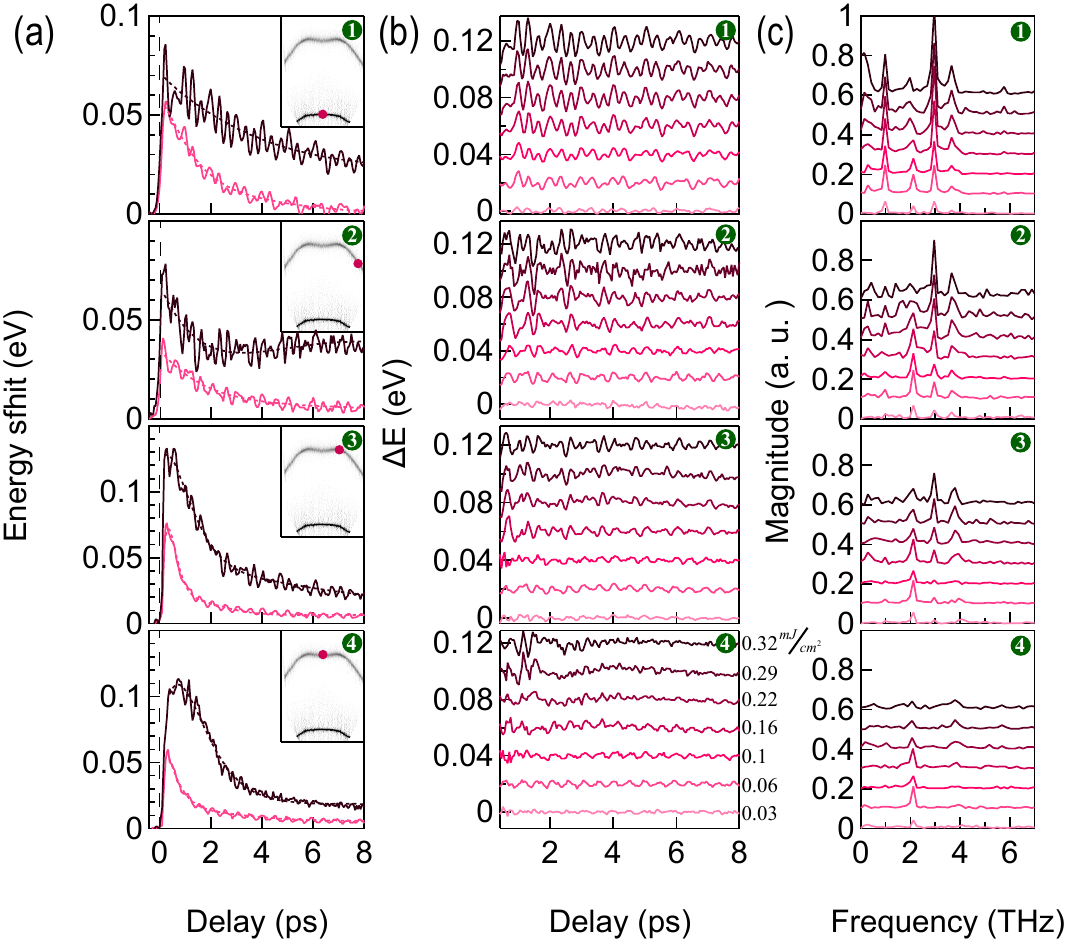}
\caption{
Time-resolved energy and momentum dependent band shifts measured at 30 K.
(a1)-(a4) Time-resolved band shifts at different energies and momentum with pump fluences of 0.1 \mJcm~ (below $F_c$) and 0.32 \mJcm~ (above $F_c$) at positions noted in the corresponding insets of each panel. The dashed lines are the incoherent part fitted to a function combined by an exponential decay function and a cubic function.
(b1)-(b4) Energy shifts after removing the incoherent part shown in panels (a1)-(a4). Similar curves measured at other four different pump fluences are shown, and the curves are offset by 0.02 eV from low to high pump fluences.
(c1)-(c4) The corresponding Fourier transform magnitudes of the curves shown in panels (b1)-(b4). Fourier transforms are done between 0.95 ps and 8 ps, and the results are offset by 0.1 from the bottom to the top.
}
\label{Fig3}
\end{figure}

%%%%%%%%%%%%%%%%%%%%%%%%%%%%%%%%%
%Fig4
%%%%%%%%%%%%%%%%%%%%%%%%%%%%%%%%%
\begin{figure}
\centering\includegraphics[width=0.8\columnwidth]{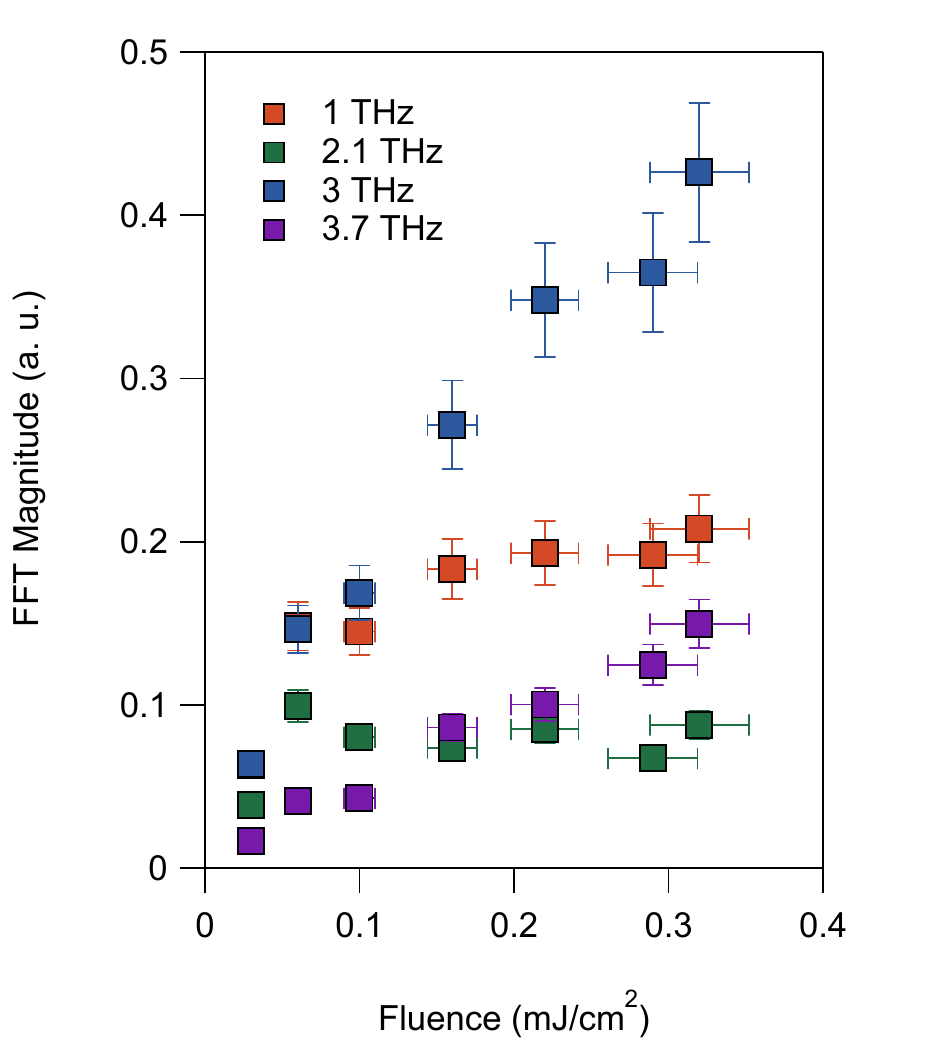}
\caption{
Pump fluence dependence of the coherent phonon vibration magnitudes from Fig.~\ref{Fig3}(c1).
}
\label{Fig4}
\end{figure}

Figure~\ref{Fig3} shows the pump-induced band shifts as a function of delay time. Interestingly, at the highest pump fluence, the pump-induced maximum shift ($\sim$ 0.13 eV) of the band \#1 top (panel (a3)) is 0.07 eV larger than that of the equilibrium shift ($\sim$ 0.06 eV) between 20 and 330 K (Fig.~\ref{Fig1} (d)), while for the band \#2 (panel (a1)) the pump-induced maximum shift ($\sim$ 0.09 eV) is 0.04 eV less than the equilibrium shift ($\sim$ 0.13 eV). The above difference suggests that the probed ultrafast dynamics is not a laser-induced heating effect. For the band \#1, the shift is larger at the momenta closer to the Fermi energy (panel (a2)-(a4)), consistent with the scenario that band \#1 is hybridized due to electronic correlations. Ultrafast band oscillations in energy are clearly identified after removing the incoherent background (panels (b1)-(b4)). At least five modes can be identified at $\sim$ 1, $\sim$ 3, $\sim$ 3.7, and $\sim$ 5.8 THz, which have A$_{1g}$ symmetry, and at 2.1 THz of the B$_{1g}$ phonon (panel (c1))\cite{Kim2016,Werdehausen2018,Larkin2018,Nakano2018,Yan2019}. All the phonon modes are shown up in the whole pump range and show little pump fluence dependence in their frequencies. The magnitudes of the 1 THz, 2.1 THz, and 3.7 THz modes are almost fluence independent while the magnitudes of the 3 THz mode is greatly enhanced with increasing the pump fluence as shown in Figure~\ref{Fig4} . However, in band \#1 the 1 THz mode is nearly absent (panel (c2)-(c4)), and at the $\Gamma$ point there is no signature of the frequency at 3 THz (panel (c4)). The absence of the 1 THz mode at the gap edge is not consistent with the scenario that this mode is strongly coupled to the excitonic condensation from ultrafast optical experiment\cite{Werdehausen2018}. The underlying mechanism of energy and momentum dependence of these coherent phonon modes is out the scope of our current study and needs to be explored theoretically in the future.

It is clear in Fig.~\ref{Fig3} that the pump-induced shifts of band \#2 oscillate synchronously exactly at the same phases and frequencies at any delay time for pump fluences both above and below the F$_c$ (panels (b1), and (c1)-(c4)). However, it has been evidenced that the phonon modes at 2.1 THz is missing above the $T_c$\cite{Mor2018}. The pump fluence independent coherent phonon modes indicates that the sample holds its low temperature monoclinic structure and does not undergo an ultrafast structural transition above the $F_c$, consistent with the ultrafast optical reflectivity experiments with much higher pump fluence\cite{Mor2018}. The observation confirms that the pump-induced ultrafast electronic phase transition in Fig.~\ref{Fig2} is purely electronic origin.

%%%%%%%%%%%%%%%%%%%%%%%%%%%%%%%%%
%Fig5
%%%%%%%%%%%%%%%%%%%%%%%%%%%%%%%%%
\begin{figure}
\centering\includegraphics[width=1\columnwidth]{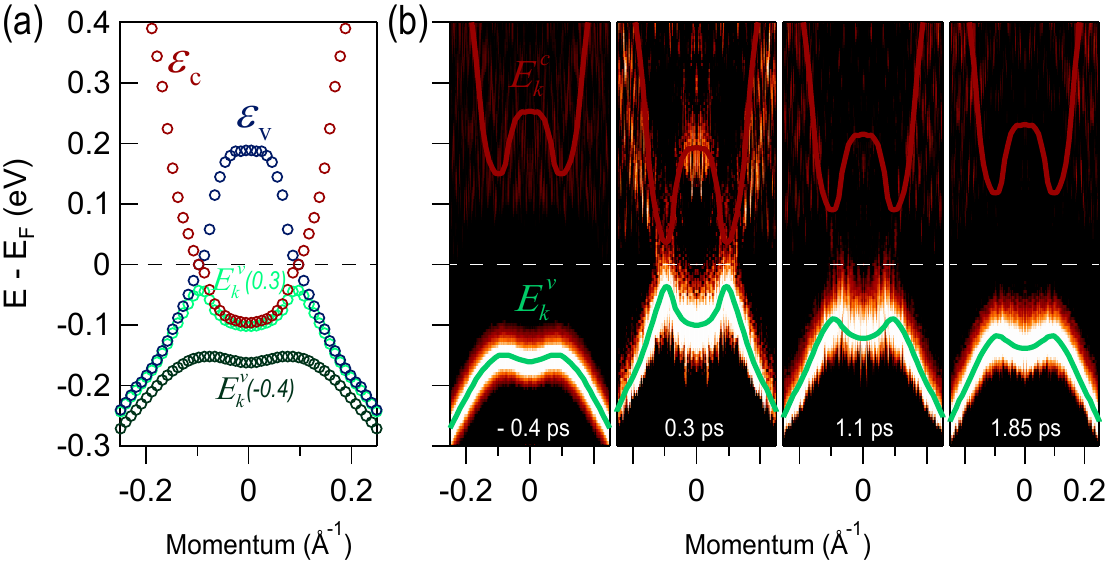}
\caption{
The bare valence and conduction bands from time-resolved electronic qusiparticle dispersions with a pump fluence of 0.19 \mJcm~ measured at 30 K.
(a) The underlying bare valence ($\varepsilon_k^v$) and conduction bands ($\varepsilon_k^c$) extracted from the experimental equilibrium dispersion $E_k^c$ at -0.4 ps and non-equilibrium dispersion $E_k^v$ at 0.3 ps.
(b) Calculated quasiparticle energies ($E_k^c(t)$ at -0.4, 0.3, 1.1, and 1.85 ps and $E_k^v(t)$) at 1.1 and 1.85 ps from the bare valence ($\varepsilon_k^v$) and conduction bands ($\varepsilon_k^c$) in panel (a) on top of the curvature images of non-equilibrium spectra.
}
\label{Fig5}
\end{figure}

In Fig.~\ref{Fig5}(a), the quasiparticle energies as a function of momentum below the Fermi energy in equilibrium state ($E_k^v(-0.4)$) at -0.4 ps and excited state ($E_k^v(0.3)$) at 0.3 ps are extracted by tracking the EDC peaks at different momentum from the spectra shown in Fig.~\ref{Fig2}(a) (Supplemental Fig. S5\cite{SM}).
From a mean field theory of the excitonic insulating physics\cite{Jerome1967,Phan2010,Yamada2016}, the quasiparticle energies of the conduction ($E_k^c$) and valence ($E_k^v$) bands are given by
\begin{equation}
\label{Ekc}
E_k^c=\frac {\varepsilon_k^c+\varepsilon_k^v}{2}+\sqrt{(\varepsilon_k^c-\varepsilon_k^v)^2/4+\Delta^2},
\end{equation}
and
\begin{equation}
\label{Ekv}
E_k^v=\frac {\varepsilon_k^c+\varepsilon_k^v}{2}-\sqrt{(\varepsilon_k^c-\varepsilon_k^v)^2/4+\Delta^2},
\end{equation}
in which $\varepsilon_k^c$ and $\varepsilon_k^v$ are the bare conduction and valence bands without electron-hole interaction, and the energy gap (order parameter $\Delta$) is the energy difference between the band top and the Fermi energy, taking the energy gap of band \#1 shown in Fig.~\ref{Fig1}(d) for an example. Since exciton is dispersionless near the $\Gamma$ point\cite{Wang2018}, it is reasonable to expect the same energy gap in the small momentum region shown in the Fig.~\ref{Fig5}. With an assumption of nearly pump independent bare dispersions, the $\varepsilon_k^c$ and $\varepsilon_k^v$ at each momenta can be obtained by solving the Eq.~\ref{Ekv} from two sets of $E_k^v$ and the corresponding energy gap at two separate delay time.% with the assumption that $E_k^v$ follows the same equation at each delay time.
The $\varepsilon_k^c$ and $\varepsilon_k^v$ in the Fig.~\ref{Fig5}(a) are the solutions of $E_k^v(-0.4)$ and $E_k^v(0.3)$.
The experimental energy gaps at delay time -0.4, 0.3, 1.1, and 1.85 ps are 0.147, 0.041, 0.118, and 0.09 eV, respectively (Supplemental Fig. S5\cite{SM}).
The obtained $\varepsilon_k^c$ and $\varepsilon_k^v$ show the semimetallic feature near the Fermi energy with a negative energy gap of about 0.285 eV, which cannot be fully captured by a tight-binding model free of excitonic interaction\cite{Sugimoto2018}. In Fig.~\ref{Fig5}(b), with the obtained bared dispersion and the energy gap at each delay time from experiments, the $E_k^c$ at each delay time and also the $E_k^v$ at 1.1, and 1.85 ps are calculated using Eqs.~\ref{Ekc} and \ref{Ekv}. Amazingly, the calculated $E_k^c$ at 0.3 ps coincides with the measured states above the Fermi energy very well and the $E_k^v$ at 1.1 ps, and 1.85 ps is consistent with the measured dispersions below the Fermi energy (panel (b) and Supplemental Fig. S5\cite{SM}), and this in turn confirms that the assumption of pump independent bare bands is reasonable here and the pump-induced chemical potential shift is negligible compare to the size of order parameter. These results show that in the momentum region probed in this experiment the quasiparticle dispersion in \TNS~can be well captured by the mean-field theory with an order parameter. As a pure electronic originated phase transition, the order parameter here is very likely related to the excitation insulating phase, since there is no other phase founded in this material. However, the obtained large order here parameter can be a combined result of pure electron-hole interaction and other strong many-body effect such as electron-phonon coupling as discussed later on.

The determined effective masses of the bare valance ($m_v$) and conduction ($m_c$) bands from Fig.~\ref{Fig5}(a) are $\sim$$0.23 m_e$ and $\sim$$0.34 m_e$, where $m_e$ is the free electron mass. Previous experiments revealed that the effective exciton binding energy $E_B$ in \TNS~at low temperature is higher than 100 meV\cite{Larkin2017}, which is reasonable . If such strong electron-hole binding is exclusively driven by electron-hole Coulomb interaction, from $E_B=\frac{m_cm_v}{2m_e(m_c+m_v)\epsilon^2}\cdot Ry$\cite{Jerome1967}, the estimated effective dielectric constant $\epsilon$ is $<$3. For a low dimensional system, the effective dielectric constant would be many times smaller that the measured, but however, the estimated dielectric constant here is more than one order smaller than the experimental result of $\thicksim$70 along the chain direction $\thicksim$40 perpendicular to the chain\cite{Larkin2018}, indicating that the strong binding in \TNS~cannot only a result of Coulomb interaction. Other many-body effects, such as electron-phonon coupling, play important roles in driving such strong electron-hole interaction in \TNS\cite{Zenker2014,Lee2019}. However, we claim that mean-field Eqs.~\ref{Ekc} and ~\ref{Ekv} are still available in a narrow momentum region near the $\Gamma$, analogous to the application of the BCS formalism in describing the superconducting electronic states of cuprates\cite{Matsui2003}.

%From the classical relation $T_c\approx0.45E_B/k_B$\cite{Jerome1967}, the estimated critical temperature at such strong electron-hole binding energy is 2150 K, which is consistent with the highest non-equilibrium electronic temperature 2008$\pm$55 K near the critical pump fluence estimated from a transient Fermi-Dirac distribution ((Supplemental Fig. S2\cite{SM})). Considering a band gap enhancement of $\sim$0.03 eV due to the orthohombic-monoclinic structural transition at the critical temperature\cite{Kaneko2013,Lee2019}, the possible semi-metallic orthohombic phase would exist still near 2000 K. Since the synthesis of \TNS~is at around 1000 K\cite{Sunshine1985}, the observed non-equilibrium semi-metallic behavior cannot be realized in the equilibrium state, indicating a hidden semi-metallic phase above the melting point in \TNS.

\section{\label{sec:level4}CONCLUSIONS}

In summary, by ultrafast photon pulse excitation, the \TNS~undergoes an ultrafast electronic phase transition with holding its monoclinic crystal structure, suggesting that in \TNS~the low temperature energy gap is a result of strong electron-hole interaction and the ultrafast phase transition is purely electronic origin. The negative gaped bare dispersions near the Fermi energy are determined experimentally. Our results and analysis suggest that the only electron-hole Columb interaction cannot be strong enough to induce the observed strong electron-hole interaction and thus many-body effects of electrons interacting with the other freedom such as lattice and orbit must be considered in further theoretical studies.
% Our studies also suggest that strong electron-electron attraction possibly sustains up to room temperature via many-body effects to realize room temperature superconductivity in material under normal condition.

\begin{acknowledgments}

We thank Z. W. Shi and W. Ku for useful discussions. W.T.Z. acknowledges support from the Ministry of Science and Technology of China (2016YFA0300501) and from National Natural Science Foundation of China (Grants No. 11674224 and 11974243) and additional support from a Shanghai talent program. T.W.T. acknowledges support from National Natural Science Foundation of China (11704247). D.Q. acknowledges support from the Ministry of Science and Technology of China (Grants No. 2016YFA0301003) and the National Natural Science Foundation of China (Grants No. U1632272, No. 11521404). Y.F.G. acknowledges the support by the Natural Science Foundation of Shanghai (Grant No. 17ZR1443300), and the National Natural Science Foundation of China (Grant No. 11874264).

\end{acknowledgments}

\end{document}

\begin{CJK*}{GBK}{}
\title{Non-Coulomb strong electron-hole binding in \TNS~revealed by time- and angle-resolved photoemission spectroscopy: Supplemental Material}

\author{Tianwei Tang}
\affiliation{Key Laboratory of Artificial Structures and Quantum Control (Ministry of Education), Shenyang National Laboratory for Materials Science, School of Physics and Astronomy, Shanghai Jiao Tong University, Shanghai 200240, China}
\author{Hongyuan Wang}
\affiliation{School of Physical Science and Technology, ShanghaiTech University, Shanghai 201210, China}
\author{Shaofeng Duan}
\author{Yuanyuan Yang}
\author{Chaozhi Huang}
\affiliation{Key Laboratory of Artificial Structures and Quantum Control (Ministry of Education), Shenyang National Laboratory for Materials Science, School of Physics and Astronomy, Shanghai Jiao Tong University, Shanghai 200240, China}
\author{Yanfeng Guo}
\affiliation{School of Physical Science and Technology, ShanghaiTech University, Shanghai 201210, China}
\author{Dong Qian}
\author{Wentao Zhang}
\email{wentaozhang@sjtu.edu.cn}
\affiliation{Key Laboratory of Artificial Structures and Quantum Control (Ministry of Education), Shenyang National Laboratory for Materials Science, School of Physics and Astronomy, Shanghai Jiao Tong University, Shanghai 200240, China}
%\date {\today}

\begin{abstract}

The material includes four supplemental discussions and five supplemental figures.
%Combined with the dramatic reduction of the scattering rate in the electron-hole mixed states below the transition temperature from high-resolution equilibrium electronic structure, these findings prove that \TNS~is an excitonic insulator.
%, and the band crossing of the valence and conduction band is slightly gaped due to the protection of the crystal symmetry, and the underlying conduction and valence bands are precisely determined from the time-resolved spectra
\end{abstract}

\maketitle
%\clearpage
\end{CJK*}
\renewcommand{\bibnumfmt}[1]{[S#1]}
\renewcommand{\citenumfont}[1]{S#1}
\makeatletter
\makeatletter \renewcommand{\fnum@figure}
{\figurename~S\thefigure}
\renewcommand{\theequation}{S\arabic{equation}}
\makeatother
%In a narrow gaped semimetal or semiconductor the bound electron-hole pairs will create a clean energy gap when the binding energy exceeds the band gap, resulting an excitonic insulating phase\cite{Mott1961,Kozlov1965,Jerome1967}. The excitonic insulating transition is purely electronic origin, and the ground state is analogous to that of the

\textbf{Supplemental discussion \#1: laser induced heating effect}

\begin{figure}%[H]
\centering\includegraphics{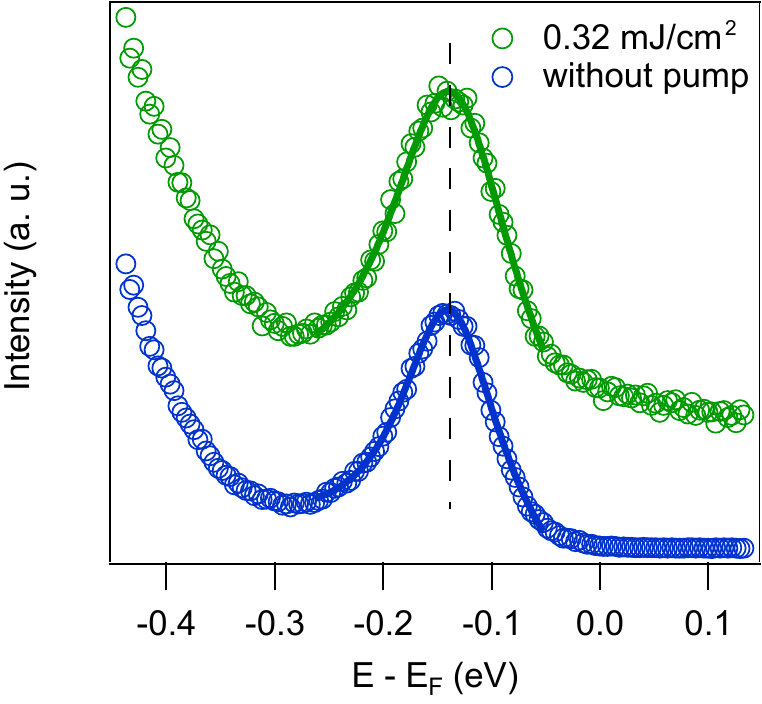}
\caption{
EDC without pump and the EDC measured at -0.4 ps with a pump fluence of 0.32\mJcm. Both of the EDCs are at the momenta of the band top (red dashed line cut in Fig. 1(a) in the main text). The sample base temperature is 30 K and the repetition rate of the laser is set to 500 kHz.
}
\label{FigS1}
\end{figure}

The repletion rate of our trARPES system is tunable by 500/n  kHz (n = 1, 2 , 3...), and we set n = 1 during the measurement. It is hard to determine the laser induced heating effect (the electronic temperature before t0) of the pump beam by simply fitting the Fermi edge because of the large energy gap at low temperature in \TNS. The cryostat we use is normally running at ~4.5 K, but to reduce the laser heating effect, we ran the experiment at 30 K. We note that the heat capacity is $\sim$ $50 Jmol^{-1}K^{-1}$ at 30 K which is tens of times larger than that at 5 K\cite{Lu2017}. In addition, we checked the gap size before $t_0$ and the equilibrium gap size measured without pump. For the highest pump fluence measured at 30 K, the energy gap before $t_0$ is ~0.142 eV, which is comparable to the equilibrium (no pump) energy gap of about ~0.144 eV at 30 K (Fig. S\ref{FigS1}). Although it is hard to determine the exact temperature before $t_0$, we can believe the temperature is far below the Tc from the temperature dependent energy gap shown in the Fig. 1(d) of the main text, and this can be further evidenced from the nearly fluence independent pump-induced coherent phonon modes shown in Fig. 3 in the main text.\\

\textbf{Supplemental discussion \#2: low energy cut-off}

\begin{figure}%[H]
\centering\includegraphics{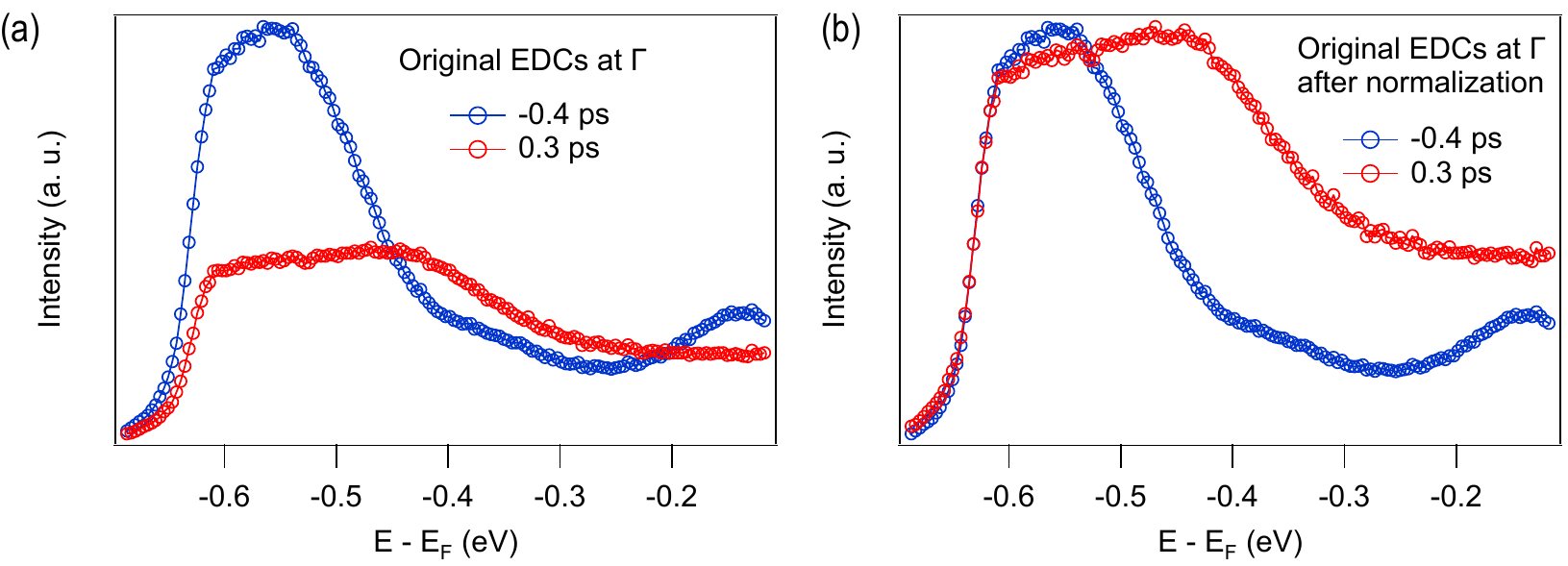}
\caption{
(a) The original EDCs at $\Gamma$ measured at 0.4 ps and 0.3 ps respectively with the highest pump fluence 0.32 \mJcm.
(b) EDCs after normalized near the low energy cut-off.
}
\label{FigS2}
\end{figure}

There is an energy of cut-off at $h\nu-\Phi$ ($\Phi$, sample work function) below the Fermi energy in photoemission. Panel (a) of Fig.~S\ref{FigS2} shows the original EDCs at $\Gamma$ measured at 0.4 ps and 0.3 ps respectively with the highest pump fluence 0.32 \mJcm. We can see that the intensity near the cut-off energy is reduced by nearly half due to the pump-induced shift of band \#2. After normalizing the height near the cut-off as show in panel (b), it is clear that there is little pump induced energy shift at this cut-off. The result shows that the photovotage effect is negligible in the measured sample up to the highest pump fluence we used.\\

\textbf{Supplemental discussion \#3: the initial decay rate}

\begin{figure}%[H]
\centering\includegraphics{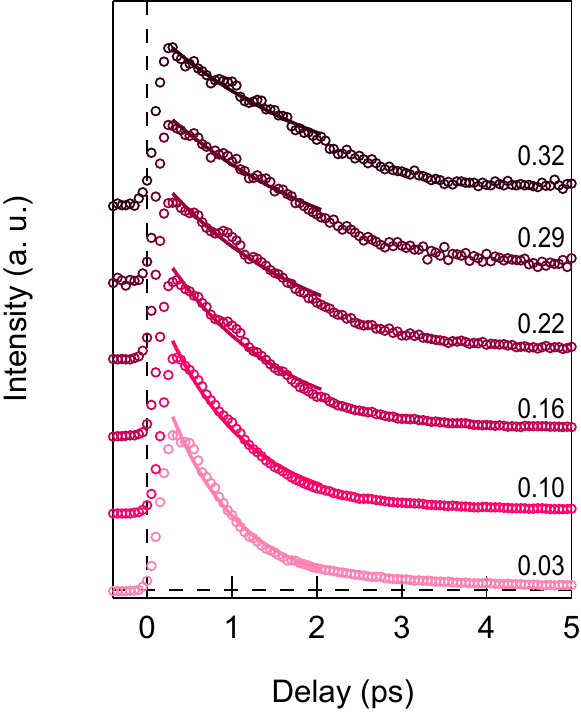}
\caption{
EDCs from Fig. 2(c) in the main text and the fittings to extract the initial decay rate of non-equilibrium electrons.
}
\label{FigS3}
\end{figure}

The initial decay rate ($\gamma$) of non-equilibrium electrons is extracted by fitting the decay curves shown in the Fig. 2(c) in the main text and also in Fig. S\ref{FigS3} to an exponential decay function (Eq.~\ref{Decay}). The decay function is an exponential decay convolved by an Gaussian resolution function $g(t)=e^{-4ln2t^2/R_t^2}$, in which $R_t$ is the time resolution of 0.13 ps here. To capture only the initial dynamics, fittings are applied on the data points between 0.3 and 2 ps.
\begin{equation}
\label{Decay}
I(t)=C_0\int e^{-t/\gamma}g(t-t')dt'
\end{equation}

\textbf{Supplemental discussion \#4: chemical potential and sample stoichiometry}

\begin{figure}
\centering\includegraphics{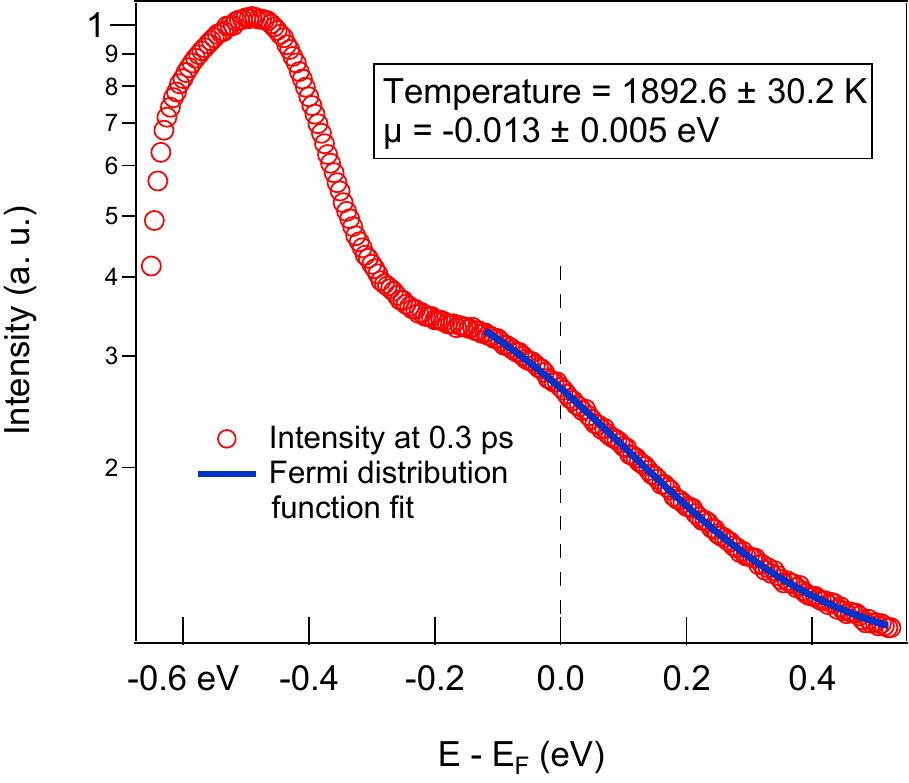}
\caption{
The non-equilibrium electronic temperature at a delay time of 0.3 ps measured with a pump fluence of 0.29 \mJcm.
}
\label{FigS4}
\end{figure}

We use an energy-resolution convolved Fermi distribution function to fit the momentum integrated intensity as a function of energy with pump fluence above the critical pump fluence in the Fig. S\ref{FigS4}. The spectrum is obtained at the delay time 0.3 ps with pump fluence 0.29 \mJcm. In the fitting, the chemical potential and electronic temperature are free parameters. The fitting result shows that the chemical potential is almost at the same energy as the Fermi energy of the ARPES system. Although it is hard to experimentally estimate the chemical potential in the gapped states below the critical pump fluence or at longer delay time, in which the electronic temperature is much lower, it is reasonable to assume the chemical potential is still near the Fermi energy.

It is hard to check the stoichiometry of the sample experimentally since the sample is always gaped up to room temperature and the conduction and valence dispersions are hard to resolve. We assume the sample we studied is near stoichiometric for the following reasons: 1) the compound was synthesized with stoichiometric ratio of elements; 2) the equilibrium energy gap 2*$\delta$ we measured is consistent with the optical and tunneling gap in similar samples\cite{Larkin2017,Lee2019}; 3) with such assumption, the solved conduction band and the top of the valence band match the experimental result after killing the gap (the spectra at 0.3 ps in Fig. 5(b) of in the main text).

\bibliography{TNS}
%%%%%%%%%%%%%%%%%%%%%%%%%%%%%%%%%
%Fig1
%%%%%%%%%%%%%%%%%%%%%%%%%%%%%%%%%

%%%%%%%%%%%%%%%%%%%%%%%%%%%%%%%%%
%FigS2
%%%%%%%%%%%%%%%%%%%%%%%%%%%%%%%%%

%%%%%%%%%%%%%%%%%%%%%%%%%%%%%%%%%
%FigS3
%%%%%%%%%%%%%%%%%%%%%%%%%%%%%%%%%
%%%%%%%%%%%%%%%%%%%%%%%%%%%%%%%%%
%FigS4
%%%%%%%%%%%%%%%%%%%%%%%%%%%%%%%%%
%\begin{figure}%[H]
%\centering\includegraphics{FigS4.pdf}
%\caption{
%Pump fluence dependence of the coherent phonon vibration magnitudes from Fig. 3(c1) in the main text.
%}
%\label{FigS4}
%\end{figure}
%%%%%%%%%%%%%%%%%%%%%%%%%%%%%%%%%
%FigS4
%%%%%%%%%%%%%%%%%%%%%%%%%%%%%%%%%
%FigS5
%%%%%%%%%%%%%%%%%%%%%%%%%%%%%%%%%
\begin{figure}
\centering\includegraphics[width=1\columnwidth]{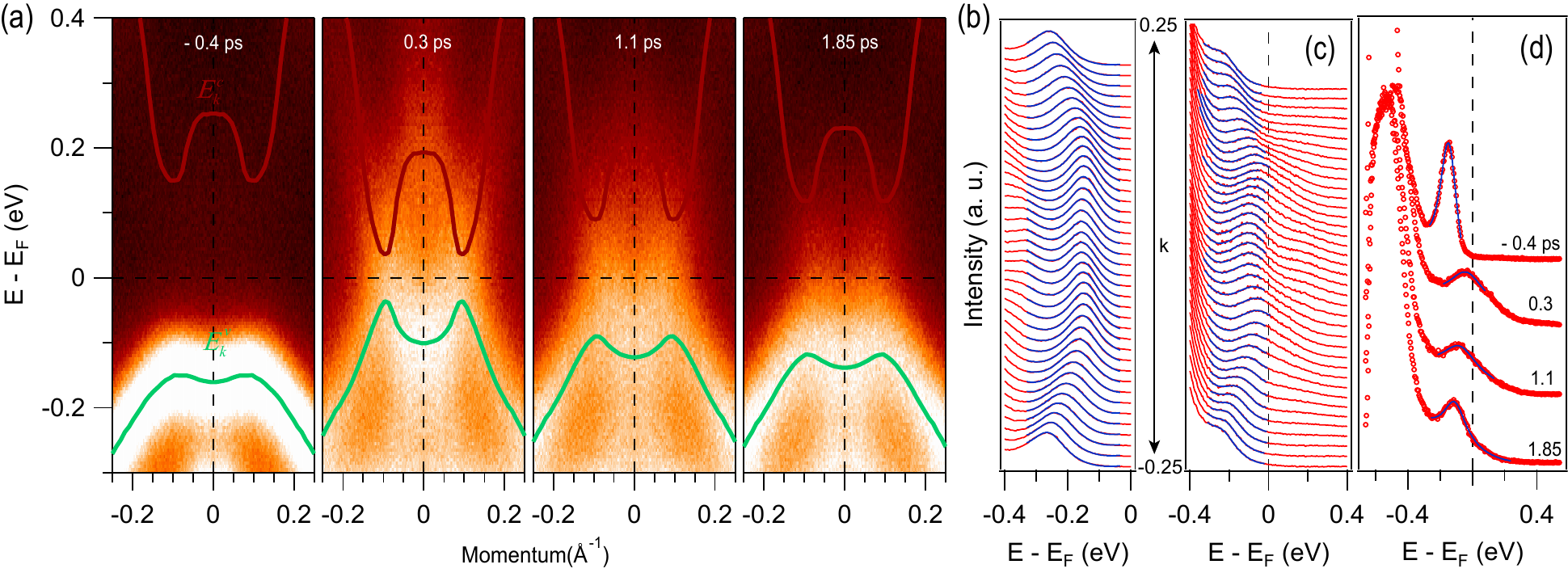}
\caption{
Time-resolved photoemission spectra with pump fluence at 0.19 \mJcm.
(a) Photoemission intensities as a function of momentum at pump-probe delay time at -0.4 ps, 0.3 ps, 1.1 ps, and 1.85 ps. The red and green curves are the same $E_k^v$ and $E_k^c$ as that shown in Fig. 4(b) in the main text.
(b) and (c), EDCs (red curves) from momentum -0.25 $\AA^{-1}$ to 0.25 $\AA^{-1}$ at the delay time -0.4 ps and 0.3 ps, respectively.
(d) EDCs at momentum 0.095 $\AA^{-1}$, i.e., the band top of $E_k^v$. The blue curves are the Lorentzian (with a linear background) fits to extract the peak positions and energy gaps.
}
\label{FigS5}
\end{figure}
%%%%%%%%%%%%%%%%%%%%%%%%%%%%%%%%%